\documentclass{fmj2010}
\usepackage{natbib} 
\usepackage{natbib} 
\bibpunct{(}{)}{;}{a}{}{,} 
\usepackage{graphicx}
\usepackage{threeparttable}

\begin{document}
   \title{Shock-Shock Interaction in the Jet of CTA\,102}

   \author{C. M. Fromm\inst{1}\thanks{Member of the International Max Planck Research School for Astronomy and Astrophysics.}, E. Ros\inst{2,1},
     T. Savolainen\inst{1}, A. P. Lobanov\inst{1}, M. Perucho\inst{2},
     J. A. Zensus\inst{1}, M. F. Aller\inst{3}, H. D. Aller\inst{3},
     M. A. Gurwell\inst{4} \and A. L\"ahteenm\"aki\inst{5} }

        \institute{Max-Planck-Institut f\"ur Radioastronomie, Auf dem
          H\"ugel 69, D-53121 Bonn, Germany 
          \and 
          Departament d'Astronomia i Astrof\'{\i}sica, Universitat de
          Val\`encia, E-46100, Burjassot, Val\`encia, Spain 
          \and 
          University of Michigan, Ann Arbor, MI, 48109 USA 
          \and
          Harvard-Smithsonian Center for Astrophysics, Cambridge, MA,
          02138 USA 
          \and 
          Aalto University, Mets\"ahovi Radio Observatory, FI-02540
          Kylm\"al\"a, Finland}

        \abstract{ The radio light curve and spectral evolution of the
          blazar CTA\,102 during its 2006 outburst can be rather well
          explained by the standard shock-in-jet model. The results of
          a pixel-to-pixel spectral analysis of multi-frequency VLBI
          images, together with kinematics derived from the MOJAVE
          survey lead to the picture of an over-pressured jet with
          respect to the ambient medium. The interaction of a
          traveling shock wave with a standing one is a possible
          scenario which could explain the observed spectral
          behavior.}

\titlerunning{The Jet in CTA\,102}
\authorrunning{C. M. Fromm et al.}
   \maketitle
%

\section{Introduction}

The blazar CTA\,102, (B2230+114), has a redshift of $z=1.037$
\citep{Hew89} and it is classified as a high polarization quasar (HPQ)
with a linear optical polarization above 3\% \citep{Ver03} with an
optical magnitude of 17.33. The source was observed for the first time
by \citet{Har60} and right from the beginning it was showing radio
variability \citep[e.g.,][]{Sho65} which led other scientists to
suggest that the signal was coming from an extraterrestrial
civilization \citep{Kar64}. Later observations identified CTA\,102 as
a quasar.

Since that time CTA\,102 has been the target for numerous observations
at different wavelengths. Besides the mentioned variation in the radio
flux density, CTA\,102 changes its optical behavior in a no less
spectacular way. \citet{Pic88} reported rapid variation up to
1.14\,mag around an average value of 17.66\,mag. An increase of
1.04\,mag within 2 days in 1978 was so far the most significant
outburst. The strongest radio flare since 1986 and a nearly
simultaneous outburst in the optical regime took place around 1997
\citep{Tor99}. Even in the $\gamma$-ray regime the source has been
detected by the telescopes EGRET on-board {\it{CGRO}} and
\textit{Fermi} with a luminosity, $L_{\gamma}=5\times10^{47}$erg/s
\citep{Nol93,Abd09}.

Within the framework of the
MOJAVE\footnote{http://www.physics.purdue.edu/MOJAVE} program
(Monitoring of Jets in Active galactic nuclei with VLBA Experiments)
CTA\,102 has been monitored since mid 1995 \citep[see][and Lister et
al. (these proceedings, p.~{159})]{Lis09b}. The results of these intensive
observations deliver a detailed picture of the morphology and
kinematics of this source. The results of the kinematic analysis show
an apparent velocity of the features in the jet between 0.7\,c and
15.40\,c \citep{Lis09b}.  A multi-frequency VLBI study including data
at $90$\,GHz, $43$\,GHz and $22$\,GHz was reported by
\citet{Ran03}. The results from the multi-frequency VLBI observations
were combined with the continuum monitoring performed at single dish
observatories at $22$\,GHz, $37$\,GHz, $90$\,GHz, and $230$\,GHz.
Within this multi-frequency data set (November 1992 until June 1998)
CTA\,102 showed a major flare around 1997. The authors could conclude
that this flare was connected to the ejection of a new jet
feature. The inferred apparent speed of $11$\,$c$, combined with the
frequent and rapid flaring events throughout the electromagnetic
spectrum, lead to the picture of a highly relativistic jet.  This
picture is supported by the conclusions of \citet{Jor05} and
\citet{Hov09}.  They found Lorentz factors, $\Gamma$, of $17$ and
$15$, and respectively, Doppler factors, $\delta$, between $15$ and
$22$.

In April 2006, CTA\,102 underwent a major radio flux outburst and we
will present the result of our analysis of this flaring event using
single dish and multi-frequency VLBI observations. We found
indications that this flaring event was created by the interaction
between a re-collimation shock and a traveling shock wave.

\section{Observations}
\subsection{Single-Dish Light Curves}

\begin{figure}{}
\resizebox{\hsize}{!}{\includegraphics{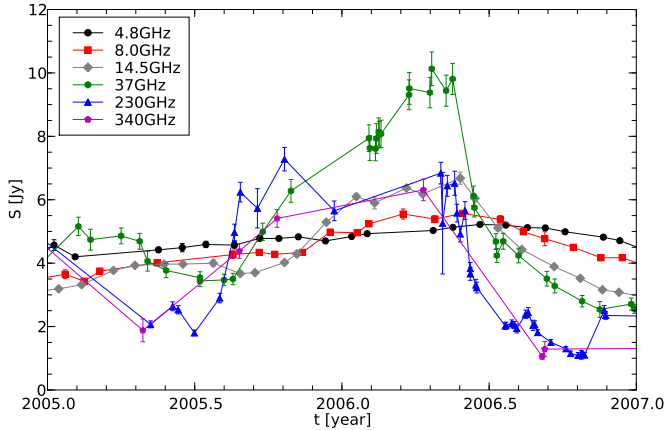}} 
\caption{Single dish light curves for CTA\,102, centered around the
  2006 radio flare. Used telescopes are 4.8--14.5\,$\mathrm{GHz}$
  UMRAO (M. F. Aller and H. D. Aller ), 37.0\,$\mathrm{GHz}$
  Mets\"ahovi (A. L\"ahteenm\"aki) and 230--340\,$\mathrm{GHz}$ SMA
  (M. Gurwell)}
\label{lightcurve} 
\end{figure}

In our analysis we concentrated on the radio flare around April 2006
and used single dish observations spanning from 4.8\,GHz to 340\,GHz
(see Fig.~\ref{lightcurve}). The flare is clearly visible at all the
frequencies with increasing time delays towards smaller
frequencies. The highest flux density of about 10\,$\mathrm{Jy}$ is
measured at 37\,$\mathrm{GHz}$.


A self-absorbed synchrotron spectrum is described by
\begin{equation}
  S(\nu)=C\left(\frac{\nu}{\nu_1}\right)^{\alpha_\mathrm{t}}\left\{1-\exp\left[-\left(\frac{\nu}{\nu_1}\right)^{\alpha_0-\alpha_\mathrm{t}}\right]\right\},
\label{snu}
\end{equation}
where $S(\nu)$ is the flux density, $\nu_1$ is the frequency at which
the opacity $\tau_s=1$, and $\alpha_\mathrm{t}$ and $\alpha_0$ are the spectral
indices for the optically thick and optically thin parts of the
spectrum, respectively. The turnover frequency, $\nu_\mathrm{m}$, and the
turnover flux density, $S_\mathrm{m}$, can be calculated from the first and the
second derivative of synchrotron spectrum and they can be regarded as
the characteristics of the spectrum.  

\begin{figure}{} 
\resizebox{\hsize}{!}{\includegraphics{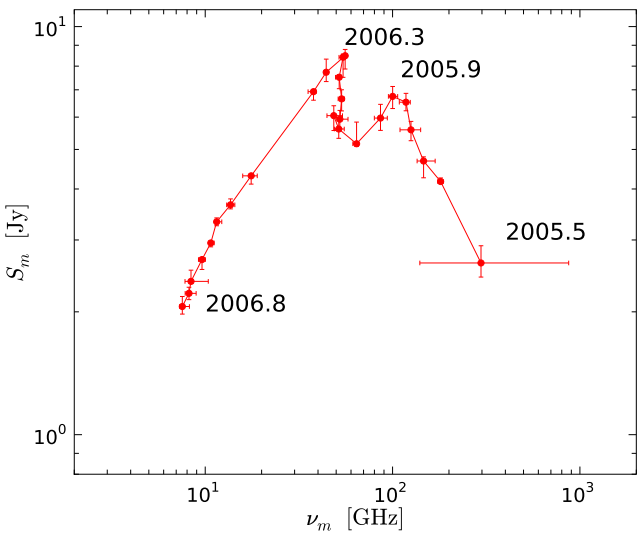}} 
\caption{The 2006 radio flare in the turnover frequency - turnover flux density plane. The numbers indicate the temporal evolution of the flare.} 
\label{numsm} 
\end{figure}

For the spectral analysis we interpolated the data to a time sampling
of $\Delta t=0.05\,\mathrm{yr}$ and subtracted an underlying quiescent
spectrum. This homogenous spectrum ($\nu_\mathrm{m}=1.70\,\mathrm{GHz}$,
$S_\mathrm{m}=4.08\,\mathrm{Jy}$, $\alpha_0=-0.35$ and $\alpha_\mathrm{t}=5/2$) was
created from archival data. The spectral evolution of the 2006 radio
flare in CTA\,102 is presented in the turnover frequency - turnover
flux density ($\nu_\mathrm{m}-S_\mathrm{m}$) plane (see Fig.~\ref{numsm}).

The flare starts around 2005.6 at a high turnover frequency
($\nu_\mathrm{m}\sim 300\,\mathrm{GHz}$) and low turnover flux density
($S_\mathrm{m}\sim 3\,\mathrm{Jy}$). During the first 0.3\,$\mathrm{yr}$ the
turnover flux density, $S_\mathrm{m}$, is increasing (to $S_\mathrm{m}\sim
6.5\,\mathrm{Jy}$) while the turnover frequency, $\nu_\mathrm{m}$ is decreasing
(to $\nu_\mathrm{m}\sim 120\,\mathrm{GHz}$). Following \cite{Mar85} we could
identify this stage as the Compton stage, where Compton losses are the
dominant energy loss mechanism. The next stage in the shock-in-jet
model should be the synchrotron one. This stage is characterized by a
less prominent changes in the turnover flux density, $S_\mathrm{m}$, while the
turnover frequency, $\nu_\mathrm{m}$ is still decreasing. One could consider
the time between 2005.8 and 2005.9 as a possible candidate for this
stage. During this time span of 0.1\,$\mathrm{yr}$ the turnover flux
density is slightly increasing (to $S_\mathrm{m}\sim 7.0\,\mathrm{Jy}$) while
the turnover frequency keeps on decreasing (to $\nu_\mathrm{m}\sim
90\,\mathrm{GHz}$). In the final stage the energy losses are dominated
by the adiabatic expansion of the jet and the relativistic shock, the
adiabatic loss stage. During this stage the turnover flux density and
and the turnover frequency are decreasing. The adiabatic losses start
to dominate the spectral evolution between 2005.9 and 2006.0.

The increase of the turnover flux density starting from 2006.0 on and
reaching a peak value of $S_\mathrm{m}\sim 8.5\,\mathrm{Jy}$ in 2006.3 can not
be explained by the shock-in-jet model. After 2006.3 the turnover flux
density decreases with decreasing turnover frequency. A detailed
analysis of the single dish observations will be presented elsewhere.

\begin{figure}{}
\centering
 \includegraphics[clip,width=\columnwidth]{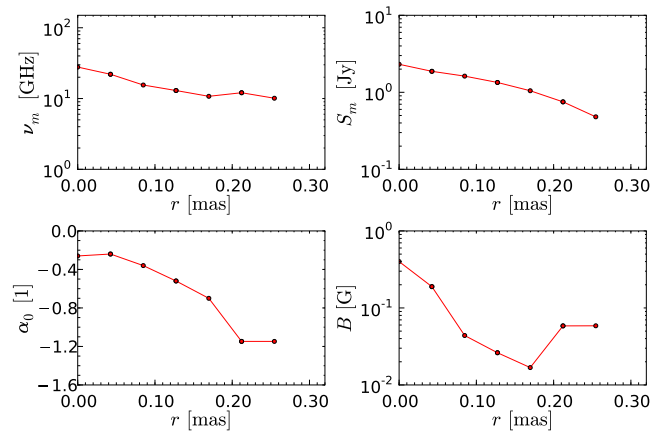}
 \caption{Evolution of the spectral parameters along the jet axis
   derived from the 2005.39 multi-frequency observations of
   CTA\,102. The estimate of the uncertainties are difficult due their
   dependence on the image alignment and the uneven uv-coverage
   between the different frequencies. Note that the individual pixels
   are not independent. An estimate for the errors, assuming correct
   alignment, on the turnover frequency and flux density is around
   15\% of the derived value (see Discussion).}
\label{specevo}
\end{figure}

\subsection {Multi-Frequency VLBI Observations}
Since single-dish observations do not provide structural information
of the jet we used as well multi-frequency VLBI observations covering
a frequency range from 2 to 86\,$\mathrm{GHz}$ for accessing the
neighborhood of the AGN central engine. CTA\,102 was observed with the
National Radio Astronomy Observatory's Very Long Baseline Array at
three epochs (March 9th 2005, April 14th 2006 and 8th June 2006) using
the all 10 antennas of the array. After calibration of the raw data,
using the standard AIPS procedures and model fitting in DIFMAP, we
performed a core-shift analysis ($\Delta r\propto \nu^{1/k_r}$)
following \citet{Lob98} and a spectral analysis on the pixels along
the jet axis using Equation \ref{snu}. From the 2005.39
multi-frequency observations of CTA\,102 we concluded from the
core-shift results that the source is in equipartition (jet particle
energy density equals magnetic field energy density). This
circumstance is reflected by a value of $k_r=0.98\pm0.03$ and leads to
an absolute distance from the central object
$r_\mathrm{core,86\,GHz}=3.29\pm1.12\,\mathrm{pc}$ and to a magnetic
field $B_\mathrm{core,86\,GHz}=0.40\pm0.14\,\mathrm{G}$. From the
derived spectral values (especially $\nu_\mathrm{m}$, $S_\mathrm{m}$)
we calculated the magnetic field $B\propto\nu_\mathrm{m}^5
S_\mathrm{m}^{-2}$.

\begin{figure}{}
\begin{centering}
\includegraphics[width=7cm]{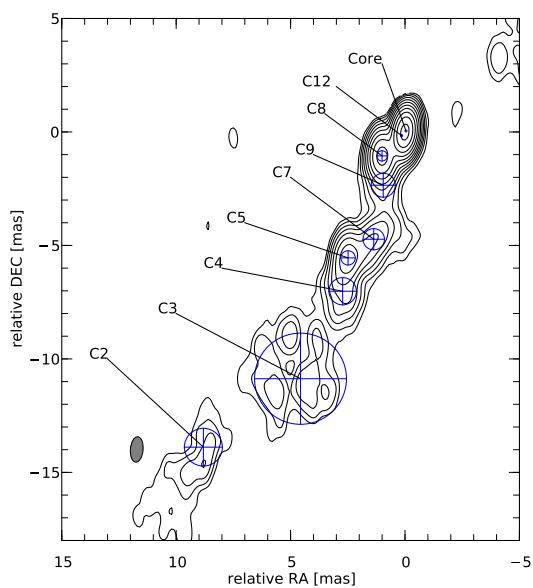}
\caption{15\,$\mathrm{GHz}$ VLBI image of CTA\,102 observed on 6th of January 2007.} 
\label{VLBI} 
\end{centering}
\end{figure}

\begin{figure}{}
\begin{centering}
\includegraphics[width=7.4cm]{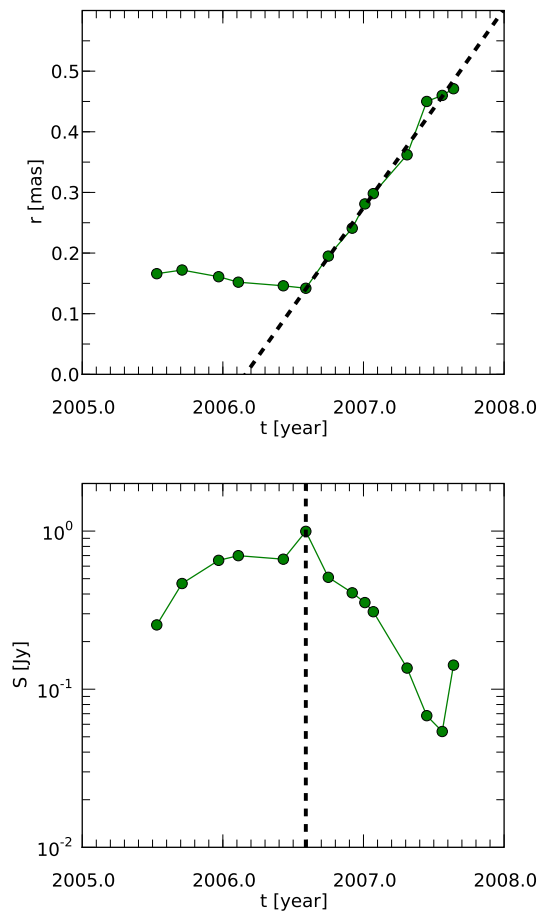}
\caption{Properties of the C12 component from the MOJAVE
  observations. Left: Separation from the core inclusive linear fit to
  the acceleration event (dashed line). Right: Evolution of the flux
  density and time of beginning of the acceleration event (dashed
  line, see text for explanation). The uncertainties for the flux
  densities are around 5\% of the total intensities and the positional
  errors are of $1/5$ of the restoring beam dimensions (typically
  around 0.01\,$\mathrm{mas}$)}
\label{MOJAVE}
\end{centering}
\end{figure}

Figure~\ref{specevo} presents the evolution of the physical parameters
deduced by our analysis as a function of the distance from the
core. The results show an increase in the turnover frequency, $\nu_\mathrm{m}$,
(upper right panel), the optically thin spectral index, $\alpha_0$,
(lower right panel) and the magnetic field, $B$, at a distance of
0.2\,$\mathrm{mas}$ away from the core. This behavior could be an
indication for an re-collimation shock (standing shock wave) at this
position.

\subsection{MOJAVE Observations}
To check the evidence for a re-collimation shock at a distance of
$0.2\,\mathrm{mas}$ from the core we used the kinematic analysis of
the 15\,$\mathrm{GHz}$ VLBI monitoring of CTA\,102 \citep{Lis09b}. A
high resolution VLBI image of CTA\,102 observed at 15\,$\mathrm{GHz}$
on January 2007, showing the location and size all fitted circular
Gaussian components, is presented in Fig.~\ref{VLBI}. The kinematics
and the evolution of the flux density of a fitted component labeled as
C12 could be interpreted as the interaction of a moving shock wave
with the re-collimation shock: the separation from the core
($r\sim0.2\,\mathrm{mas}$) remains constant until mid 2006, when an
acceleration event takes place together with a sharp drop in the flux
density (see Fig.~\ref{MOJAVE}).

By fitting the acceleration part of the C12 trajectory and correcting
for the core-shift we derived a value for the ejection time,
$t_\mathrm{ej}=2005.83\pm0.05\,\mathrm{yr}$. This time corresponds to
the first peak in the $\nu_\mathrm{m}$-$S_\mathrm{m}$ plane. The second peak is located
slightly before the time of the beginning of the acceleration event
(possible intersection of the moving shock wave with the
re-collimation shock)
$t_\mathrm{intersec}\sim2006.6\,\mathrm{yr}$. Using a viewing angle of
$\vartheta=2.6\pm0.5^\cdot$ \citep{Jor05} we derived a velocity
$\beta=v/c=0.998\pm0.042$ and a bulk Lorentz factor
$\Gamma=17.7\pm0.7$.

\section{Discussion}
The evolution of the 2006 radio flare in the $\nu_\mathrm{m}$-$S_\mathrm{m}$ plane could
be explained by the interaction of a moving shock wave with a
re-collimation shock. Re-collimation shocks are stationary features in
non-pressure matched jets and lead to local increase in pressure and
change in the orientation and value of the magnetic field
\citep{Dal88,Fal91,Per07}. This behavior can be seen in the increase
of the turnover frequency, $\nu_\mathrm{m}$ and the increase in the magnetic
field, $B$ at a distance of 0.2\,$\mathrm{mas}$ in the evolution of
the spectral parameters derived from multi-frequency VLBI observations
(see Fig.~\ref{specevo}). The first peak in the $\nu_\mathrm{m}$-$S_\mathrm{m}$ plane
corresponds to the ejection of the traveling shock wave around
$t_\mathrm{ej}\sim2005.85\,\mathrm{yr}$. The trajectory of this new
feature can not be observed at 15\,$\mathrm{GHz}$ due to the limited
resolution. The interaction of the moving shock wave with
re-collimation shock takes place around mid 2006 which could lead to
the second peak in the $\nu_\mathrm{m}$-$S_\mathrm{m}$ due to shock acceleration in a
region of increased pressure and magnetic field. During the
interaction of the two waves the re-collimation shock is pulled away
by the moving one and should appear after some time again at the same
position. One possible explanation why this reaction is not detected
in the 15\,$\mathrm{GHz}$ kinematics could be a rarefaction wave. This
wave is traveling in the wake of the moving shock wave and decreases
the pressure behind the shock front. Together with the limited
dynamical range of the VLBI this could be the reason for the
non-detection of the re-collimation shock after the collision with the
traveling one.
 
\section{Conclusions and Outlook}
The combination of single-dish observations with multi-frequency,
densely time-sampled VLBI monitoring is a powerful approach which can
contribute towards a better understanding of flaring events. Using a
high-quality set of observational data we presented a possible
scenario which could explain the observed flare by a shock-shock
interaction and derived estimates for the physical parameters of the
jet and the traveling shock wave.

The spectral values presented have been derived from a spectral
analysis applied on multi-frequency VLBI observations. This technique
is sensitive to the correct image alignment and to effects of the
uneven uv-coverage between the frequencies. The effect on the derived
parameters due to small misalignments could lead to significant
changes in the spectral values, especially in the magnetic field
($B\propto\nu_\mathrm{m}^5 S_\mathrm{m}^2$). Further analysis of the
influence of the alignment is needed to provide adequate error bars
for the derived values.

Besides quantifying the uncertainties on the observational parameters
we started to test our assumption of the shock-shock interaction using
2D relativistic magneto-hydrodynamic simulations. These simulations
will help us to understand the formation of re-collimation shocks in
magnetized jets and their interactions with traveling shocks.

It is also of interest to investigate possible correlations between
$\gamma$-ray flares and the collision of re-collimation shocks and
traveling shock waves. These investigations could help to clarify the
question where in the jet the high energy radiation is generated.

\begin{acknowledgements}
  We thank C.S.~Chang for valuable comments and inspiring
  discussions. CMF  was supported for this research through a
  stipend from the International Max Planck Research School (IMPRS)
  for Astronomy and Astrophysics. 
  M. Perucho acknowledges support from a ``Juan de la
  Cierva'' contract of the Spanish ``Ministerio de Ciencia y
  Tecnolog\'{\i}a'', the Spanish ``Ministerio de Educaci\'on y
  Ciencia'' and the European Fund for Regional Development through
  grants AYA2007-67627-C03-01 and AYA2007-67752-C03-02 and
  Consolider-Ingenio 2010, ref. 20811. This work is based on
  observations with the radio telescope of the university of Michigan,
  MI, USA, the Mets\"ahovi radio telescope of the university of
  Helsinki, Finland and Sub-Millimeter Array (SMA) of the Smithsonian
  Astrophysical Observatory, Cambridge, MA, USA. The operation of
  UMRAO is made possible by funds from the NSF and from the university
  of Michigan. The Submillimeter Array is a joint project between the
  Smithsonian Astrophysical Observatory and the Academia Sinica
  Institute of Astronomy and Astrophysics and is funded by the
  Smithsonian Institution and the Academia Sinica. The Mets\"ahovi
  team acknowledges the support from the Academy of Finland to our
  observing projects
\end{acknowledgements}

\bibliography{biblio}

\end{document}